\documentclass[aps,prl,twocolumn,showcaps,superscriptaddress,showpacs,amsmath,amssymb,floatfix]{revtex4}

\usepackage{graphicx}
\usepackage{dcolumn}
\usepackage{bm}
\usepackage[dvips]{epsfig}

\usepackage{color}

\newcommand {\dr}{{\mathrm d}\mathbf{r}}

\begin{document}

\title{Criticality and phase separation in a two-dimensional binary colloidal fluid induced by the solvent critical behavior}

\author{O.~Zvyagolskaya}
\affiliation{2. Physikalisches Institut, Universit\"at Stuttgart,
Pfaffenwaldring 57, 70569 Stuttgart, Germany}

\author{A.J.~Archer}
\affiliation{Department of Mathematical Sciences, Loughborough University, Loughborough, LE11 3TU, United Kingdom}

\author{C.~Bechinger}
\affiliation{2. Physikalisches Institut, Universit\"at Stuttgart,
Pfaffenwaldring 57, 70569 Stuttgart, Germany}
\affiliation{Max-Planck-Institut f\"ur Intelligente Systeme,
Heisenbergstr.3, 70569 Stuttgart, Germany}

\begin{abstract}
We present an experimental and theoretical study of the phase behavior of a binary mixture of colloids with opposite adsorption preferences in a critical solvent. As a result of the attractive  and repulsive critical Casimir forces, the critical fluctuations of the solvent lead to a further critical point in the colloidal system, i.e. to a critical colloidal-liquid--colloidal-liquid demixing phase transition which is controlled by the solvent temperature. Our experimental findings are in good agreement with calculations based on { a simple approximation for the free energy of the system}.

\end{abstract}

\maketitle

The structure, phase behavior and dynamics of colloidal suspensions is largely determined by the interactions between the colloids and considerable effort goes into controlling the pair potentials in such systems. This can be achieved, e.g.\ by adding salt to the solution in order to screen electrostatic interactions \cite{BH_book}, by adsorbing charged or polymeric species onto the surfaces of the colloids \cite{BH_book}, or by adding a depletion agent to the solution, such as a non-adsorbing polymer \cite{BH_book,Poon02}. While these methods have been successfully employed in many experiments, they can not be used when colloidal interactions have to be changed in-situ and in a reversible manner. In such situations, external fields may be used and have the additional advantage of allowing for time-dependent control of the colloidal interactions. Examples are superparamagnetic or dielectric colloidal particles, where the structure and phase transitions can be easily controlled by external magnetic and electric fields \cite{Hoffmannetal06,Yethiraj2003} or temperature-responsive microgel colloidal particles, which change their size when the temperature is varied by several degrees \cite{thermoresponsive}.

Only recently, critical Casimir forces have been demonstrated to provide an alternative approach for the in-situ control of colloidal forces \cite{soyka2008, Hertlein08, nellen2009, BCPP10}. Such interactions arise between particles that are immersed in a fluid mixture which is near to its critical point \cite{fisher1978}. The confinement of the critical concentration fluctuations between the surfaces of a pair of colloids leads to a force whose {range and amplitude strongly depend} on the correlation length in the mixture and thus shows a strong temperature dependence \cite{burkhardt1995, schlesener2003}. In addition, critical Casimir forces can be changed from attractive to repulsive by changing the adsorption preference of the colloidal particles for the mixture's components, which determines the corresponding boundary conditions (BC) \cite{nellen2009}. So far, all experiments have been performed with colloidal systems where all particles have the same BC which then leads to particle aggregation or glass formation in concentrated suspensions \cite{soyka2008,Beysens:JSP99, lu2010, Bonn2009}.

In this Letter, we investigate the phase behavior of a binary colloidal mixture with opposite boundary conditions that are dispersed in a binary critical mixture. Due to the attractive and repulsive critical Casimir forces for particles with the same and opposite BC respectively, we observe a critical colloidal-liquid--colloidal-liquid (CL-CL) demixing phase transition which is entirely driven by the critical concentration fluctuations of the solvent; i.e.\ the critical point in the solvent mixture {\em generates} the phase transition in the colloidal system. This demonstrates a strong coupling of different critical phenomena over more than two decades in length scales. Our measurements are in agreement with the theoretically calculated phase diagram obtained {from a simple approximation for the free energy of the system}.

\begin{figure}
  \includegraphics[width=7.5cm]{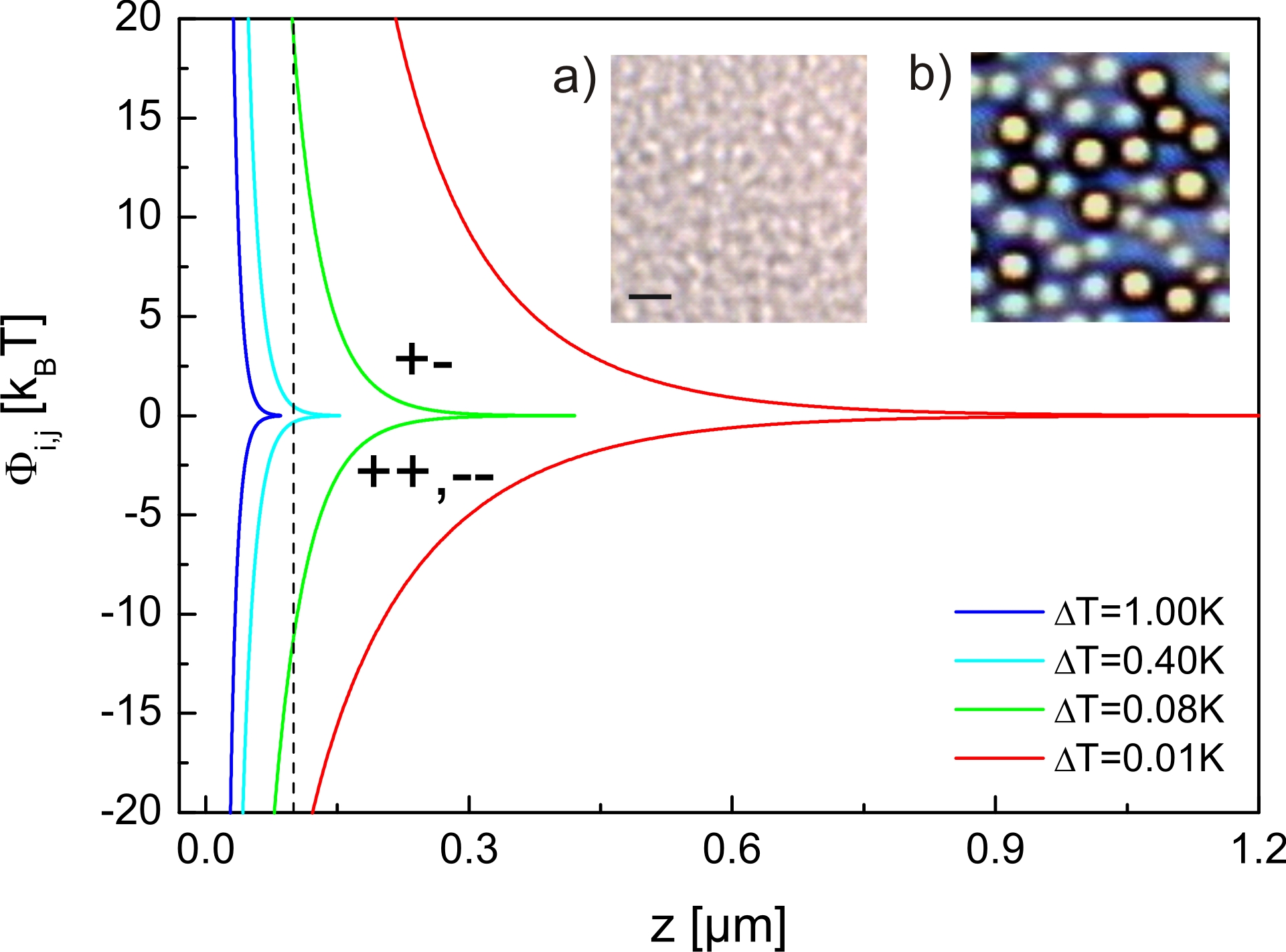}
  \caption{Critical Casimir potentials obtained from Eq.\ \eqref{eq:1} with the scaling functions from Ref.\ \cite{Hertlein08}, for two $3\,\mu$m sized particles with symmetric $(+ +, - -)$ and antisymmetric $(+ -)$ boundary conditions at various different temperatures. Inset: a) Experimental snapshot of the demixing process of the critical water -- 2,6-lutidine mixture. b) Snapshot of the binary colloidal mixture, with one species labeled with a fluorescent dye. The scale bar corresponds to $3\,\mu $m.}
   \label{setup}
\end{figure}
\begin{figure*}
 \includegraphics[width=17cm]{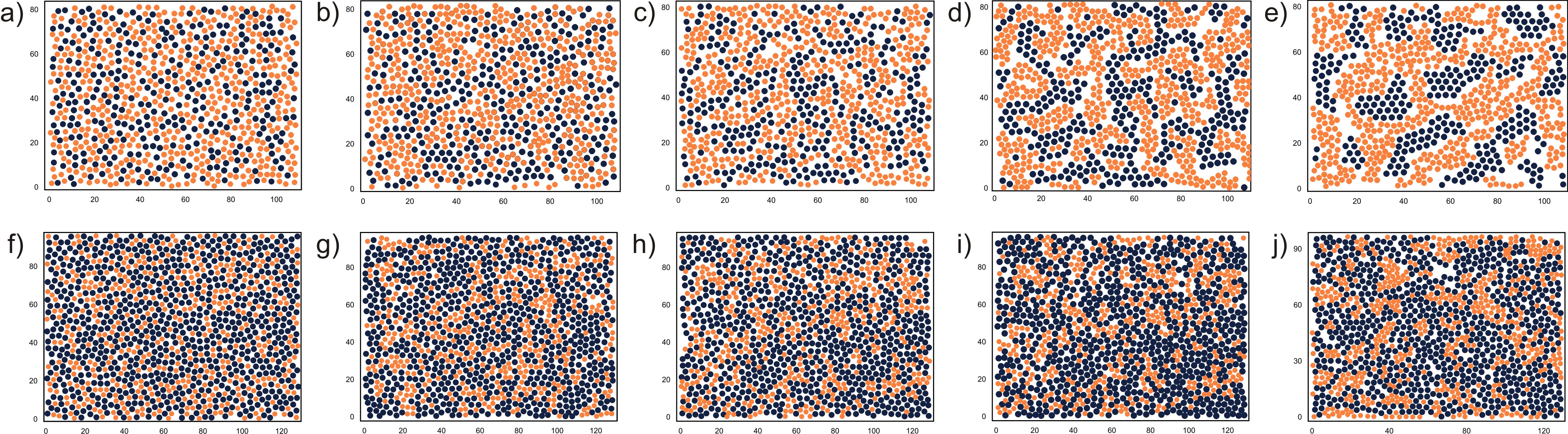}
    \caption{Snapshots of the particles in binary colloidal mixtures of two different compositions: in panels a-e, $x_A=0.54$ and in f-j, $x_A=0.29$. The temperature is approaching $T_c$, as one goes from a to e and also from f to j, respectively. The last image in each column is a snapshot of the system at a longer time. Panels a-d are obtained after 1h at the temperatures $\Delta T=1$K, 0.40K, 0.08K and { 0.01K} and e is obtained after 20h at $\Delta T=$0.01K. Panels f-i are obtained after 1h at $\Delta T=1$K, 0.09K, 0.05K and 0.02K and j is obtained after 22h at $\Delta T=$0.02K.}
  \label{result_1}
\end{figure*}

The critical Casimir potential between two colloidal spheres with radius $R$ with center-center distance $r$ is given by \cite{Hertlein08}:
\begin{equation}
\frac{ \Phi_{ij}(z)}{k_BT}=\frac{R}{2z}\theta_{ij}\left(\frac{z}{\xi}\right),
\label{eq:1}
\end{equation}
where $z=r-2R$ is the distance between the particle's surfaces and $\xi=\xi_0 (\Delta T/T_c)^{-0.63}$ is the solvent bulk correlation length of the binary mixture with a critical temperature $T_c$. The value $\xi_0$ is a microscopic length scale which characterizes the range of the molecular interactions in the mixture. The function $\theta_{ij}(z/\xi)$ is a universal scaling function with ($i=j$) corresponding to symmetric and  ($i\neq j$) antisymmetric BC. Note the factor $1/2$ in Eq.\ \eqref{eq:1}, since we used the scaling function for a sphere-wall geometry \cite{schlesener2003}.

Our experimental system is composed of a binary mixture of silica spheres with slightly different diameters $\sigma_A=2.8\,\mu$m and $\sigma_B=3.4\,\mu$m which are suspended in a water-2,6-lutidine mixture at critical composition (lutidine mass fraction of $c^c_L\cong 0.286$). The solvent is characterized by a lower critical temperature $T_c=307$K (i.e. the mixture is in its homogeneous state below $T_c$) and $\xi_0 \approx 0.2$nm$\pm0.02$nm, as determined from light scattering experiments \cite{guelari1972}. Initially, both species of particles (denoted $A$ and $B$) have a strong adsorption preference for water, i.e.\ with $(-)$ BC, due to their high surface charge \cite{GALLAGHER1992a}. In order to reverse the BC of the A-type particles, they are functionalized with silane according to a procedure described in \cite{kamal}. {This treatment renders the particles hydrophobic, i.e. (+) BC, as inferred from the fact that they preferentially stay in the lutidine-rich phase of the demixed binary solvent.} The entire colloidal suspension is contained in a temperature-controlled thin sample cell in which {all} the colloids sediment towards the bottom plate and form a rather dense monolayer. The sample plane is imaged onto a camera and the particle positions are determined with video microscopy. To facilitate the discrimination of particles with opposite BC, the B-type particles are labeled with a fluorescent dye (see inset of Fig.$\,$\ref{setup}). We should mention that for the particle size ratio used here, no demixing due to depletion interactions has been observed. {We should also mention that critical Casimir forces also arise between the particles and the bottom of the container. However, these do not influence the observed demixing because these forces only change the average particle height above the substrate and so, e.g., change the particle mobilities due to hydrodynamic interactions \cite{Brenner,BP00}. In case of the silica particles used here, such variations in hight can not be experimentally resolved due to the large gravitational forces which localize the particles at roughly the same hight above the surface. As a result, the change in the mobility of the particles is negligible. Note also that bridging \cite{Hertlein08,HAE09} between the particles and the substrate does not occur because we use a mixture with a critical composition and always stay below the critical temperature.}

Fig.$\,$\ref{setup} shows the critical Casimir interaction $\Phi_{ij}(r)$ for different temperatures between particles with 3 $\mu m$ diameter and with symmetric $(++, - -)$ and antisymmetric $(+ -)$ BC as calculated from  Eq.\ \eqref{eq:1}. Additionally, electrostatic contributions have to be considered, which depend on the particle's surface charge and the Debye screening length of the solvent with  $\kappa^{-1}\approx 12$nm for a critical water-lutidine mixture  \cite{Hertlein08}. Because the position of the first peak in the pair correlation {functions of the colloidal mixture are independent of the particle density in the temperature range covered in our experiments, the electrostatic interactions are well approximated by a hard sphere potential with an effective particle diameter $\sigma_{eff}=\sigma_i+0.1\mu $m (dashed vertical line in Fig.$\,$\ref{setup}).}

\begin{figure}
 \includegraphics[width=6cm]{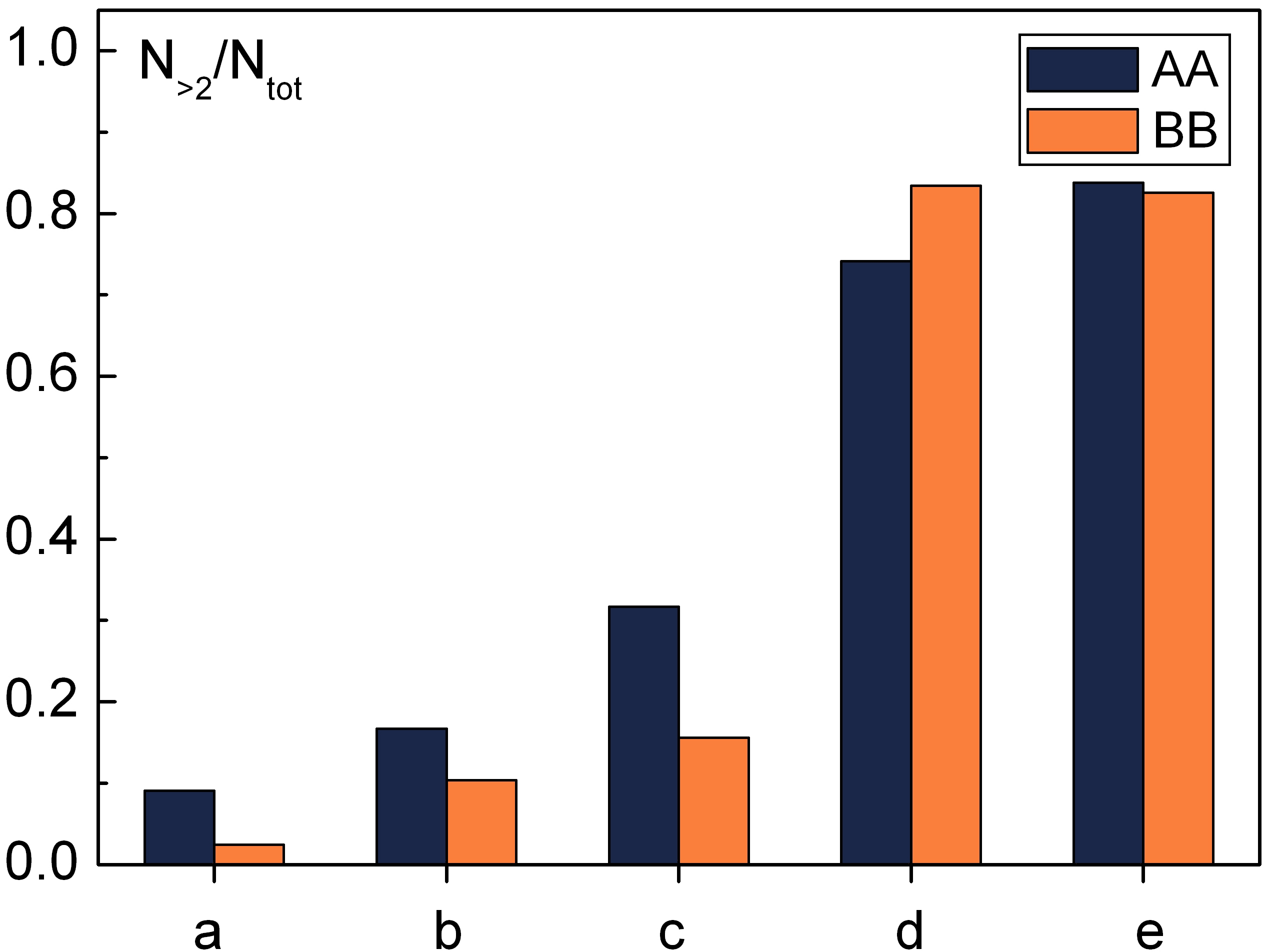}
 \caption{Fraction of particles $N_{c>2}/N_{tot}$ with coordination number $c>2$, where $N_{tot}$ is the total number of particles, calculated separately for each kind of particles (AA and BB), for the snapshots in Figs.\ \ref{result_1}a-e.}
 \label{fig5}
 \end{figure}

Fig.~\ref{result_1} shows temperature dependent configurations of a binary colloidal system with total density {$\rho_A \sigma{_A}{^2} + \rho_B \sigma{_B}{^2} =0.70 \pm 0.05$}, where $\rho_A$ and $\rho_B$ are the (two-dimensional) surface  {number} densities of the $A$ (orange) and the $B$ (black) particles for two mixing ratios $x_A=\rho_A/(\rho_A+\rho_B)= 0.54$ (Figs.~\ref{result_1}$\,$a-e) and $x_A=0.28$ (Figs.~\ref{result_1}$\,$f-j). Each temperature change is followed by a waiting time of about 1h to allow the particles to rearrange accordingly. When increasing $T$ from room temperature towards $T_c$, up to $\Delta T = 0.4K$, the particles remain homogeneously distributed for both compositions (Figs.$\,$\ref{result_1}$\,$a,b). This clearly suggests that in this temperature range, the {critical Casimir pair interaction energy is less than} or equal to the thermal energy as also confirmed by Fig.$\,$\ref{setup}. Further approaching the critical point, the critical Casimir forces increase and we observe large structural changes in the colloidal mixture. For $x_A=0.54$ the system changes from a random distribution to an almost bicontinuous network. {We see this beginning to set in, in Fig.$\,$\ref{result_1}$\,$c, which corresponds to $\Delta T = 0.08K$}. Further increasing the temperature of the solvent to $\Delta T = 0.01 K$ leads to a coarsening of the structure (Fig.$\,$\ref{result_1}d). This coarsening continues as a function of time, when the sample is kept at the same temperature. This is demonstrated by Fig.\ \ref{result_1}e which shows the same situation after a waiting time of a 20h. {The process of the phase separation taking place in Figs.\ \ref{result_1}a-e can be followed more quantitatively by determining the number and type of the nearest neighbors for each particle, which we define as the number of particles with centers a distance $<1.1\sigma_i$ away from a given particle of the same type (A,B). In Fig.\ \ref{fig5} we display the fraction of each type of particles with a coordination number $c>2$. We see that the fraction is low when the particles are homogeneously distributed (Figs.\ref{result_1}a-b), but grows rapidly to a value close to 1 when the system is phase separated due to the fact that in this situation it is more likely for particles to be surrounded by other particles  of the same species (Figs.\ \ref{result_1}d-e). The coordination number $c=2$ was the most frequently found configuration in the snapshot Fig.\ref{result_1}c, for both kinds of particles, with $N_{c=2}/N_{tot}$ ranging from $0.3$ to $0.4$. This coordination number mostly corresponds to string-like colloidal structures which we believe are the initial stages of phase separation, forming the beginnings of a bicontinuous network.}

\begin{figure*}
 \includegraphics[width=11cm]{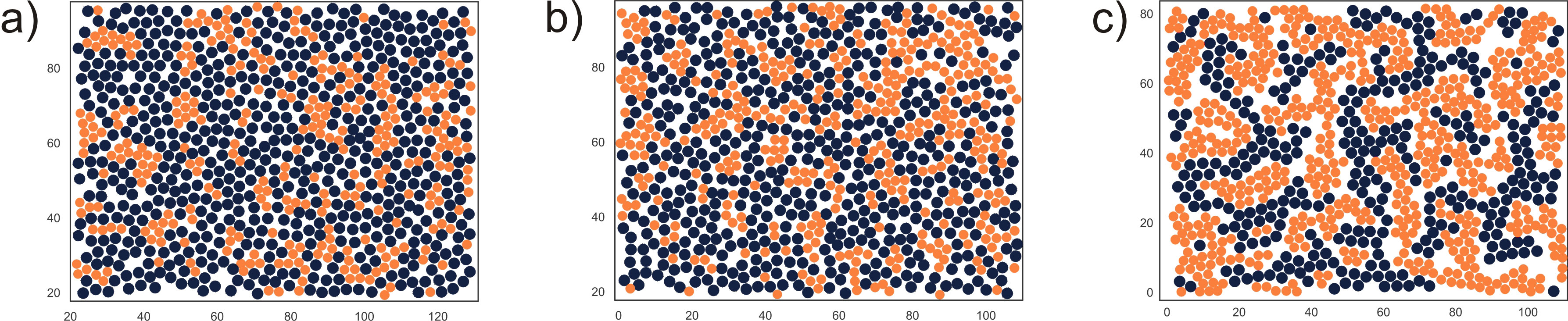}
    \caption{Snapshots of binary colloidal systems of three different compositions, $x_A=0.28, x_A=0.32$ and $x_A=0.54$, obtained after 1h, at temperatures close to $T_c$, with $\Delta T=$0.01K.}
  \label{result_2}
\end{figure*}

For $x_A=0.28$ the corresponding experiments do not show the formation of a bicontinuous network and the CL-CL demixing instead proceeds via the growth of small clusters of the minority phase ($A$ particles) which are surrounded by a liquid composed of the majority ($B$-type) particles. Similar to the case described above, the arrangement displayed in Fig.\ \ref{result_1}i (formed after $\approx1$h) hardly changes when the sample is kept at the same temperature for a further 20h (Fig.$\,$\ref{result_1}$\,$j) {as the coarsening is rather slow}. Such a strong dependence of the morphology of the demixing process on the particle composition has been already observed by other authors \cite{RD89}. Here, however,  the CL-CL demixing transition and associated critical point is induced by {\em another} critical point (that of the solvent), which demonstrates a coupling of critical phenomena over many decades in length scales. Additional evidence for the nature of the CL-CL demixing process may be obtained from Fig.\ \ref{result_2}, where we show snapshots for various different compositions at almost identical values of $\Delta T$. We observe a change from the `island' morphology at lower values of $x_A$ to a bicontinuous structure when $x_A\approx0.5$.

{Having established from the experimental images (e.g.\ by comparing Figs.\ \ref{result_1}a and \ref{result_1}d) that approaching the solvent critical temperature leads to colloidal phase separation, we now develop a model for the system free energy in order to further understand the phase behavior of the system and to construct a phase diagram.} For simplicity, we consider a mono-disperse, two-dimensional systems with the following particle interactions:
\begin{equation}
u_{ij}(r) =
\begin{cases}
\infty \hspace{17mm} r \leq \sigma_{eff} \\
\Phi_{ij}(r) \hspace{11mm} r > \sigma_{eff},
\end{cases}
\label{eq:pair_pot}
\end{equation}
where the indices $i,j=A,B$ and $\Phi_{ij}(r)$ is the corresponding critical Casimir potential \eqref{eq:1} acting between the colloidal particles (we use the universal scaling functions $\theta_{ij}(z/\xi)$ from Ref.\,\cite{Hertlein08}) and the electrostatic {repulsion is modelled via a hard disk repulsion with effective diameter $\sigma_{eff}=3.1\mu$m. For the other parameters we use the corresponding experimental values (see above).}

We construct an approximation for the Helmholtz free energy of the system by splitting the pair potential into the hard-disk part $u_{hd}(r)$ plus the tail term:
\begin{equation}
w_{ij}(r) =
\begin{cases}
0 \hspace{18mm} r \leq {\sigma_{eff}} \\
\Phi_{ij} (r) \hspace{10mm} r > {\sigma_{eff}}.
\end{cases}
\label{eq:pair_pot_2}
\end{equation}
Thus, the full pair potential between the particles is $u_{ij}(r)=u_{hd}(r)+w_{ij}(r)$. {We assume the following mean-field (Van der Waals like) approximation for the Helmholtz free energy per unit area:
\begin{eqnarray}
f(\rho_A,\rho_B)=f_{hd}(\rho_A,\rho_B)+\frac{1}{2}\sum_{ij} \rho_i \rho_j \hat{w}_{ij},
\label{eq:F}
\end{eqnarray}
where $\hat{w}_{ij}=\int \dr \,w_{ij}(r)$ and $f_{hd}$ is the Helmholtz free energy per unit area of a uniform fluid of hard-disks with bulk densities $\rho_A$ and $\rho_B$. We have used the scaled particle approximation for $f_{hd}$ -- see Ref.\ \cite{Rosenfeld}. $f_{hd}$ includes the (exact) ideal gas contribution $f_{id}=k_BT\sum_i \rho_i(\ln[\Lambda^2 \rho_i]-1)$, where $k_B$ is Boltzmann's constant and $\Lambda$ is the thermal de Broglie wavelength.}

\begin{figure}
  \includegraphics[width=8cm]{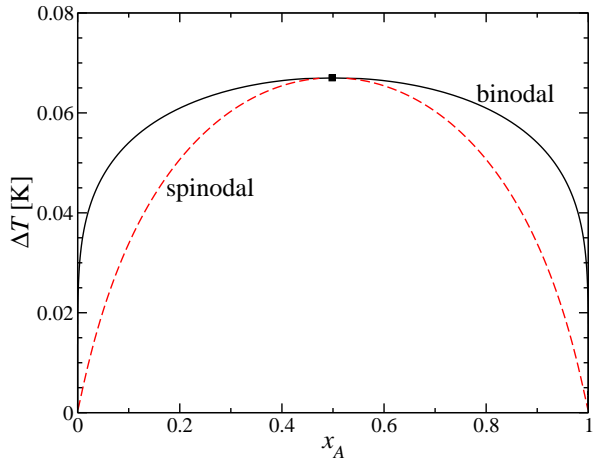}
  \caption{Phase diagram for the system plotted as a function of $\Delta T$ versus concentration $x_A=\rho_A/(\rho_A+\rho_B)$, for colloids with a total density {$\rho_A\sigma_A^2+\rho_B\sigma_B^2=0.7$}. The index $A$ and $B$ correspond to the $(+)$ and $(-)$ particles in the snapshots displayed in Figs.\ \ref{result_1} and \ref{result_2}. We see that the theory predicts that for $\Delta T<{0.067}$K the Casimir forces between the colloids are strong enough to induce phase separation.}
  \label{phasediag}
\end{figure}

As the temperature of the solvent approaches $T \to T_c$, the {strength} of the Casimir potential increases and eventually leads to a CL-CL demixing transition. Within the simple model described above, the phase behavior depends on the values of the integrals {$\hat{w}_{ij}$}, and therefore on $\Delta T$. In Fig.\ \ref{phasediag} we display the resulting phase diagram {for the case when the total density $\rho_A\sigma_A^2+\rho_B\sigma_B^2=0.7$, which corresponds to the experimental results in Figs.\ \ref{result_1} and \ref{result_2}.} Because the strength of the critical Casimir forces are identical among like species of particles, this results in a symmetric demixing line (binodal) centered on the mixing ratio $x_A=0.5$. We see from Fig.\ \ref{phasediag} that as $\Delta T$ is decreased, if the mixture has the composition $x_A=0.5$, then one hits the critical point for the CL-CL demixing transition at $\Delta T \approx {0.067}K$. We also see that the onset of demixing for a range of values of $x_A$ centered on $x_A=0.5$ occurs at a similar value of $\Delta T$. This is in fairly good agreement with the experiments for $x_A=0.54$: looking at Fig.\ \ref{result_1} one would estimate that c), which is for $\Delta T=0.08$K, is near the onset of demixing.

{The position of the critical point in Fig.\ \ref{phasediag} is fairly sensitive to the size of the electrostatic contribution to $\sigma_{eff}$. Bearing in mind also some of the simplifications in our model, e.g.\ that we assumed the same size of both particle types, the difference between the experimentally determined CL-CL demixing point from the theoretically calculated value is surprisingly small. Therefore, it is clear that the results in Fig.\ \ref{result_1}c, whilst not being exactly at the critical value of $\Delta T$, are certainly for a state point nearby.}

One issue that we should comment on further is that, as mentioned above, the demixed particle configurations of our two-dimensional fluid change very slowly over time -- compare, for example, the snapshots in Fig.\ \ref{result_1}d and e, which are taken after 1h and 20h, respectively. This is due to that fact that the characteristic length scale $L(t)$ of the domains is a power-law function of time $L(t)\sim t^n$, where we expect the exponent $n\approx1/3$ \cite{Bray89,Bray90}. This means that to observe further coarsening in the size of the domains, one would have to wait over much longer time periods. Furthermore, the dynamics is also slow due to the critical slowing down that one observes in the vicinity of a critical point. We have used dynamical density functional theory (DDFT) \cite{MT1,Archer7}, based on the free energy functional {which for a uniform fluid gives the free energy} in Eq.\ \eqref{eq:F}, in order to study the demixing dynamics as the binary colloidal fluid is quenched into the spinodal region of Fig.\ \ref{phasediag}. {We see qualitative agreement between the DDFT results (not displayed) and the snapshots from the experiments, displayed in Figs.\ \ref{result_1} and \ref{result_2}. However, matching the DDFT to the experimental snapshots is not straightforward because an input to the DDFT is the long time diffusion coefficients $D_A$ and $D_B$ of the particles over the surface. One can estimate $L(t)$ from a time sequence of experimental images, allowing us to estimate $D_A\approx D_B=D$ and then one finds that the DDFT density profiles have roughly the same $L(t)$ and appear qualitatively similar to the experimental snapshots.}

In this Letter we have presented experimental and theoretical results demonstrating that one can use critical Casimir forces to precisely control the phase behavior, structure and dynamics of binary mixtures of colloidal particles. In particular, we have observed that by taking the solvent temperature close to $T_c$, one can generate a further critical point in the binary colloidal mixture. This second critical point and the associated CL-CL demixing transition is generated entirely by the solvent critical fluctuations and resulting Casimir forces due to the temperature proximity to $T_c$. This striking observation also leads one to speculate that it would be interesting to make a system to  which are added two additional species of much larger colloids. The Casimir force between these larger colloids, due to the critical fluctuations of the smaller colloids, {could then induce a third critical point in the system, although, of course, depletion forces would also be important in such a system.}

\begin{acknowledgments}
We thank Roland Roth
for helpful discussions. We also thank Kamalakannan Kailasam and Klaus M\"uller for preparing the silan grafted particles. CB and AJA gratefully acknowledge support by the Deutsche Forschungsgemeinschaft (BE 1788/{9}) and from RCUK.

\end{acknowledgments}

\bibliographystyle{unsrt}

\end{document}